\documentclass[12pt]{article}%
\usepackage[nosort]{cite}
\usepackage{graphicx}
\usepackage{subcaption}
\usepackage{multicol}
\usepackage{amsfonts}
\usepackage{amssymb}
\usepackage{amsmath}
\usepackage{heck}
\usepackage{setspace}
\usepackage{verbatim}
\usepackage{color}
\usepackage{longtable}
\usepackage{float}
\usepackage{epsfig}
\usepackage{epstopdf}
\usepackage{tikz}
\usepackage[margin=1in]{geometry}
\usepackage{titletoc}%
\providecommand{\U}[1]{\protect\rule{.1in}{.1in}}
\pdfoutput=1
\usetikzlibrary{decorations.markings}

\numberwithin{equation}{section}

\newcommand{\ba}{\begin{eqnarray}}
\newcommand{\ea}{\end{eqnarray}}

\newcommand{\decoup}{\sqcup}
\begin{document}

\date{June 2015}

\title{Evidence for C-theorems in 6D SCFTs}

\institution{UNC}{\centerline{${}^{1}$Department of Physics, University of North Carolina, Chapel Hill, NC 27599, USA}}

\institution{CUNY}{\centerline{${}^{2}$CUNY Graduate Center, Initiative for the Theoretical Sciences, New York, NY 10016, USA}}

\institution{HARVARD}{\centerline{${}^{3}$Jefferson Physical Laboratory, Harvard University, Cambridge, MA 02138, USA}}

\authors{Jonathan J. Heckman\worksat{\UNC, \CUNY}\footnote{e-mail: {\tt jheckman@email.unc.edu}} and
Tom Rudelius\worksat{\HARVARD}\footnote{e-mail: {\tt rudelius@physics.harvard.edu}}}

\abstract{Using the recently established classification of 6D SCFTs we present evidence
for the existence of families of weak C-functions, that is, quantities which decrease in a flow
from the UV to the IR. Introducing a background R-symmetry
field strength $R$, and a non-trivial tangent bundle $T$ on the 6D spacetime,
we consider C-functions given by the linear combinations $C = m_1 \alpha + m_2 \beta + m_3 \gamma$,
where the $\alpha_i$ are the anomaly polynomial
coefficients for the formal characteristic classes $c_2(R)^2$, $c_2(R) p_1(T)$ and $p_1(T)^2$.
By performing a detailed sweep over many theories,
we determine the shape of the unbounded region in ``m-space'' compatible
with both Higgs branch flows and tensor branch flows. We also verify that
--as expected-- the Euler density conformal anomaly
falls in the admissible region.}

\maketitle

\tableofcontents

\enlargethispage{\baselineskip}

\setcounter{tocdepth}{2}

\newpage

\section{Introduction \label{sec:INTRO}}

There is a basic intuition in quantum field theory that renormalization group
flows lead to a decrease in the number of degrees of freedom in a system. For
conformal field theories (CFTs), this can be made quantitative if a weak
C-function exists:\footnote{There is also a notion of a \textquotedblleft
strong C-function\textquotedblright\ which is about monotonicity of a quantity
along an entire RG flow.}%
\begin{equation}
C_{UV}\geq C_{IR}.
\end{equation}
For even-dimensional CFTs on a curved background, Cardy proposed in reference
\cite{Cardy:1988cwa} that the quantity $a_{D}$ which appears in the trace
anomaly (see e.g. \cite{Capper:1974ic}):%
\begin{equation}
\left\langle T_{\mu}^{\mu}\right\rangle =-\left(  -\frac{1}{4\pi}\right)
^{D/2}a_{D}E_{D}+...
\end{equation}
provides such a C-function. Here, $E_{D}$ is the D-dimensional Euler density
constructed with respect to the background metric, and the other contributions
to the anomaly are constructed from the Weyl tensor, as well as (scheme
dependent) divergences of currents. In the case of odd-dimensional CFTs, there
is a related quantity given by the universal part of the free energy. A proof
of the 2D a-theorem (i.e. the c-theorem) was given by Zamolodchikov in
reference \cite{Zamolodchikov:1986gt}, and a proof of the 4D a-theorem was
given recently in reference \cite{Komargodski:2011vj}. For 3D systems, there
is the related F-theorem \cite{Myers:2010tj, Jafferis:2010un, Casini:2011kv}.
For recent work on the status of the 6D a-theorem, see e.g.
\cite{Elvang:2012st} and \cite{Grinstein:2015ina}. See also
\cite{Cordova:2015vwa} for a recent proof for flows between $(2,0)$ SCFTs, and
see \cite{Cordova:2015fha} for its status in the case $(1,0)$ SCFTs.

Our aim in this note will be to study the existence of weak C-functions for
6D\ SCFTs. We use the recently completed classification of 6D\ SCFTs (see
references \cite{Heckman:2013pva, Gaiotto:2014lca, DelZotto:2014hpa,
Heckman:2014qba, DelZotto:2014fia, Heckman:2015bfa, Bhardwaj:2015xxa} as well
as earlier work \cite{Witten:1995ex, Witten:1995zh, Strominger:1995ac,
WittenSmall, Ganor:1996mu,MorrisonVafaII,Seiberg:1996vs, Seiberg:1996qx,
Bershadsky:1996nu, Brunner:1997gf, Blum:1997fw, Aspinwall:1997ye,
Intriligator:1997dh, Hanany:1997gh}) which can be generated by
compactifications of F-theory to search for C-functions which obey the
inequality $C_{UV}>C_{IR}$. Since the R-current and stress tensor
sit in the same Weyl multiplet \cite{Howe:1983fr,
Bergshoeff:1985mz}, the conformal anomalies of a 6D\ SCFT are given by
specific linear combinations of coefficients appearing in the anomaly
polynomial:%
\begin{equation}
\mathcal{I}=\alpha c_{2}(R)^{2}+\beta c_{2}(R)p_{1}(T)+\gamma%
p_{1}(T)^{2}+\delta p_{2}(T)+...
\end{equation}
Here, we have introduced formal characteristic classes $c_{2}(R)$ for the
R-symmetry, and $p_{1}(T)$ for a tangent bundle for a formal eight manifold.
The \textquotedblleft...\textquotedblright\ refers to additional theory
specific flavor symmetry field strengths. A concrete algorithm for computing
the anomaly polynomial of a 6D\ SCFT was obtained in references
\cite{Ohmori:2014pca, Ohmori:2014kda}
(see also \cite{Intriligator:2014eaa, Heckman:2015ola} as well as \cite{Freed:1998tg, Harvey:1998bx}).

We seek to determine possible four-component vectors $\overrightarrow{m}$ such
that the resulting linear combination:%
\begin{equation}
C=\overrightarrow{m}\cdot\overrightarrow{\alpha}%
\end{equation}
satisfies $C_{UV}>C_{IR}$. Here, we have introduced the four-component vector
$\overrightarrow{\alpha}=(\alpha,\beta,\gamma,\delta)$. In
lower-dimensional systems, the existence of non-trivial dualities leads to
tight bounds on possible C-functions. For example, even before the 4D
a-theorem was proved, it was already known that $a_{4D}$ was the only
candidate C-function which could be constructed from a linear combination of
the anomaly polynomial coefficients \cite{Anselmi:1997ys}. We find, however,
that in six dimensions there exist entire families of weak C-functions. This
is in part due to the fact that for our 6D theories, there do not appear to be
non-trivial dualities.

From this perspective, it is natural to determine the precise contours of the region
of ``m-space" that decreases monotonically under RG flows. First of all, we can already see
on general grounds that the coefficient $\delta$ is the same in the
UV$\ $and IR \cite{Heckman:2015ola, Cordova:2015fha},
and so $m_{4}$ is a \textquotedblleft
null\textquotedblright\ direction. For Higgs branch flows, $m_{3}$ is also a
\textquotedblleft null\textquotedblright\ direction and $m_{1}>0$. In the case
of tensor branch flows, we also find the general analytic bound $m_{1}%
m_{3}>m_{2}^{2}$.

We expect that the more complex a SCFT is, the weaker the expected bound. Though we have not performed a systematic sweep over the
more complex cases of reference \cite{Heckman:2015bfa}, we have found that in all of these cases, there is a strictly bigger
jump in the behavior of a candidate C-function compared with less complex theories. For this reason, we
focus on all theories with a single tensor multiplet, as well as the
\textquotedblleft classical theories\textquotedblright\ of reference
\cite{Heckman:2015bfa}. A classical theory has a tensor branch in which only
classical gauge groups appear. Though we have not performed a systematic sweep over theories with generalized
conformal matter \cite{DelZotto:2014hpa, Heckman:2014qba}, we expect these cases to provide weaker
bounds. The reason is that a conformal matter system is already an interacting
SCFT, and therefore contains many degrees of freedom. Higgsing such a theory
is therefore expected to generate a jump which is bigger compared with
Higgsing a classical theory.

We find that in the case of Higgs branch flows, the
tightest constraint comes from the flow of the rank 
one E-string theory to a set of free hypermultiplets. 
This leads us to the bound:%
\begin{align}
\text{Higgs\,\,(Numeric)} &  \text{: }0<m_{1}-\frac{11m_{2}}{26}.
\label{higgsbound1}
\end{align}
In the case of tensor branch flows, we find that the tightest bounds typically
come from theories with a single $SO$ or $Sp$ gauge group factor, although
theories with multiple gauge groups also make an appearance. In fact, we find
that if we formally continue the anomaly polynomial for the single $SO$ and $Sp$ case to a
rank which is a general rational number, we can cleanly state all of the
resulting bounds in terms of the behavior of a single anomaly polynomial. This
leads us to the bounds:%
\begin{subequations}
\begin{align}
\text{Tensor\,\,(Numeric)} &  \text{:\ }0<m_{1}-\frac{m_{2}}{2}+\frac{m_{3}}{16} \label{tensor1}\\
\text{Tensor\,\,(Numeric)} &  \text{: }0<m_{1}+\frac{m_{2}}{6}+\frac{m_{3}}{144} \label{tensor2}\\
\text{Tensor\,\,(Analytic)} & \text{: }0<m_{1}  \label{tensor3}\\
\text{Tensor\,\,(Analytic)} &  \text{: }m_{2}^{2}<m_{1}m_{3}.  \label{tensor4}
\end{align}
The bounds (\ref{higgsbound1}), (\ref{tensor1}), and (\ref{tensor2}), and (\ref{tensor3}) are necessary conditions: any four-vector $\overrightarrow{m}$ that fails to satisfy these conditions will not lead to a monotonically decreasing C-function.  The bound (\ref{tensor4}) on the other hand, is a sufficient condition: any $\overrightarrow{m}$ satisfying this bound will lead to a C-function that is monotonically decreasing under \emph{tensor branch} flows.  However, we will see strong evidence that the converse statements are also true in the appropriate regime: for suitably small $|m_2/ m_1|$, the four necessary conditions (\ref{higgsbound1} - \ref{tensor3}) are sufficient, whereas for sufficiently large $|m_2 / m_1|$, the sufficient condition (\ref{tensor4}) is more or less necessary.

As a simple application of our analysis, we present strong evidence that the
specific combination found in reference \cite{Cordova:2015fha} (see also \cite{EHOLOG}):
\end{subequations}
\begin{subequations}
\begin{equation}
a_{6D}=\frac{8}{3}(\alpha-\beta+\gamma)+\delta
\end{equation}
is indeed well within the monotonic region of m-space. Indeed, a proof of
this fact for tensor branch flows was also presented in \cite{Cordova:2015fha}, but as far as
we are aware no proof is yet available for Higgs branch flows. Let us also
emphasize that even if an a-theorem is eventually established through some
extension of the methods found in \cite{Cordova:2015fha}, it is still important to note that
we seem to have whole families of weak C-functions in six dimensions.

The rest of this note is organized as follows. First, in section
\ref{sec:FLOWS} we collect some general details on the empirically observed
behavior of the anomaly polynomial under RG flows. After this, we turn in section \ref{sec:ANALYSIS}
to our determination of m-space for 6D\ SCFTs. We also discuss the specific case of
the conformal anomaly $a_{6D}$. We conclude in section \ref{sec:CONC}.

\section{RG\ Flows and Anomaly Polynomials \label{sec:FLOWS}}

In this section we discuss in general terms the class of RG\ flows we shall
consider, as well as what we know about changes in the anomaly polynomial in
flowing from the UV to the IR. Our discussion mainly follows that given in
reference \cite{Heckman:2015ola}.

We focus on RG flows which correspond to geometric deformations of an F-theory
compactification. Recall that in F-theory, we generate 6D\ SCFTs by working
with a singular elliptically fibered Calabi-Yau threefold $X\rightarrow B$
with base $B$. We reach an SCFT when two-cycles of the base $B$ simultaneously
contract to zero size.

The data of the base of an F-theory compactification is conveniently
summarized by a writing down a diagram of $\mathbb{P}^{1}$'s with prescribed
self-intersection numbers. These numbers can range from $1$ up to $12$ (see \cite{Morrison:2012np} for details),
and must obey the condition that minus the intersection pairing defines a positive
definite quadratic form (see \cite{Heckman:2013pva}). Here is an example of a base configuration of curves:%
\end{subequations}
\begin{equation}
\text{Example of a Base:\ }\underset{N}{\underbrace{2,2....2,2}}.
\end{equation}
When the elliptic fiber is trivial and all of the $-2$ curves of this
configuration collapse to zero size, we realize the $A_{N}$ $(2,0)$ SCFT. In
addition to specifying the base, it often happens that the elliptic fiber is
non-trivial. As explained in \cite{Morrison:2012np}, this occurs generically
for curves with self-intersection $-n$ for
$3\leq n\leq12$. It can also happen for $-1$ and $-2$ curves by further
tuning the complex structure moduli of an F-theory model. We can incorporate
this additional data by the notation:%
\begin{equation}
\underset{[N_f = k]}{\overset{\mathfrak{g}}{n}},
\end{equation}
that is, we have a curve of self-intersection $-n$ wrapped by a seven-brane
with gauge symmetry algebra $\mathfrak{g}$. Finally, the number of flavors $[N_f = k]$ which
is often dictated by anomaly cancellation considerations has also been indicated. Unless otherwise indicated,
a flavor will refer to a hypermultiplet in the fundamental representation (as this is by far the most common situation).
We shall denote such flavor contributions by square brackets. Indeed, since we do not
actually need the flavor symmetry group, we shall find it more convenient to
simply list all the flavors. Here is an example of a $(1,0)$ SCFT which
exhibits all of the rules mentioned above:%
\begin{equation}
\text{Example }(1,0)\text{ SCFT: \ \ }[N_{f}%
=k]\underset{N}{\underbrace{\overset{\mathfrak{su}_{k}}{2}%
,\overset{\mathfrak{su}_{k}}{2}....\overset{\mathfrak{su}_{k}}{2}%
,\overset{\mathfrak{su}_{k}}{2}}}[N_{f}=k]
\end{equation}

The classification results of references  \cite{Heckman:2013pva,
Gaiotto:2014lca, DelZotto:2014hpa, Heckman:2014qba, DelZotto:2014fia,
Heckman:2015bfa, Bhardwaj:2015xxa} amount to a determination of all possible
bases and all possible fiber decorations which can occur. An important outcome
of this analysis is that all of these theories can be viewed as generalized
quivers in which the links (i.e. matter) between distinct gauge group can
sometimes also be interacting fixed points \cite{DelZotto:2014hpa, Heckman:2014qba}.
Such links therefore have many interacting degrees of freedom. That means Higgs or tensor
branch flows involving conformal matter will likely proceed by larger jumps
in a candidate weak C-function and will therefore likely give weaker bounds.

Having given a brief review of some aspects of how to build 6D\ SCFTs, let us
now turn to RG flows for these theories. There are two qualitatively distinct
kinds of RG\ flows which can be seen as deformations of the F-theory geometry.
First of all, there are tensor branch flows. Geometrically these correspond to
resolving some of the singular curves of the base $B$. In the 6D effective
field theory, this corresponds to giving a vev to the scalar of a tensor
multiplet. In this flow, the UV R-symmetry remains unbroken (as the scalar is
neutral under $SU(2)_{R}$). Second of all, there are Higgs branch flows.
Geometrically these correspond to a complex structure deformation of the
6D\ SCFT. All of these flows can be thought of as being triggered by a vev for
generalized conformal matter (in the sense of \cite{DelZotto:2014hpa,
Heckman:2014qba}). In these flows, the UV R-symmetry is broken, but an
R-symmetry is recovered in the infrared.

Now, as empirically observed in reference \cite{Heckman:2015ola}, the change
in the anomaly polynomial for tensor branch flows and Higgs branch flows is:%
\begin{align}
\Delta_{\text{Tensor}}\mathcal{I}  &  =(pc_{2}(R)-qp_{1}(T))^{2}\\
\Delta_{\text{Higgs}}\mathcal{I}  &  =c_{2}(R)(rc_{2}(R)-sp_{1}(T)).
\end{align}
Dirac quantization of the lattice of string charges enforces the condition
that $p,q,r,s$ are rational numbers \cite{Intriligator:2014eaa, Maxfield:2012aw}.
Additionally, in the case of tensor branch flows, the
factorization into a perfect square is required to be consistent with general
't Hooft anomaly matching considerations \cite{Intriligator:2014eaa}. Indeed,
since this involves introducing an additional tensor multiplet which can then
be exchanged to cancel off this contribution, the overall sign of the
difference is also fixed to be positive. In the case of the Higgs branch
flows, the fact that the UV R-symmetry is broken means that if we switch off
$c_{2}(R)$, the change in the anomaly polynomial should vanish. This enforces
the general form found here. Note that in both cases, there is no contribution
from $p_{2}(T)$ because diffeomorphisms remain unbroken along the entire flow
\cite{Heckman:2015ola}.

Phrased in this way, our task reduces to sweeping over all possible flows for
6D\ SCFTs, determining admissible values of $p,q,r$ and $s$. Though we do not
have a general proof, we have found that in all known flows, we have:
\begin{equation}
r,s>0.
\end{equation}
Returning to our general expression for our candidate C-function
$C=\overrightarrow{m}\cdot\overrightarrow{\alpha}$, we can now calculate the
change under both tensor branch and Higgs branch flows:%
\begin{align}
\Delta_{\text{Tensor}}C &  =m_{1}p^{2}-2m_{2}pq+m_{3}q^{2}\label{tensorflows}%
\\
\Delta_{\text{Higgs}}C &  =m_{1}r-m_{2}s.
\end{align}
Without specifying too many details, we can now establish two rather crude
bounds on the monotonic region of m-space. First of all, we observe that from the structure of
Higgs branch flows, $m_{1}>0$. Second of all, from the structure of tensor
branch flows, we see that our expression in $p$ and $q$ defines a quadric in
$p$ and $q$. The condition that this is positive imposes the condition that
the discriminant $\mathcal{D}$ $=4m_{2}{}^{2}-4m_{1}m_{3}$ is strictly
negative. Putting this together, we can already obtain an analytic cut through
m-space:%
\begin{align}
0  & <m_{1}\\
m_{2}{}^{2}  & <m_{1}m_{3}%
\end{align}

While it is in principle possible to proceed further by purely analytic means,
we shall instead resort to the explicit classification of 6D\ SCFTs to start
selecting candidate C-functions. The rest of this section is organized
as follows. First, we give some examples of RG\ flows, illustrating the
general point that we expect the tightest bounds to come from the simplest
theories. Next, we turn to the explicit class of theories for which we perform our sweeps. We then
turn to a summary of our results from a numerical sweep over possible theories.

\subsection{Example Flows and ``Bogus Theories''}

Some examples of tensor branch and Higgs branch flows were considered in
\cite{Heckman:2015bfa} and \cite{Heckman:2015ola}. The simplest Higgs branch
flows start with a UV theory of a single tensor multiplet paired with a vector
multiplet and appropriate hypermultiplet matter and flow to an IR theory of at most one tensor multiplet and a different gauge symmetry, along with the appropriate
hypermultiplet matter. An example is the flow,
\[
\underset{\lbrack N_{f}=1]}{\overset{\mathfrak{e_{6}}}{5}}%
\overset{RG}{\rightarrow}\overset{\mathfrak{f_{4}}}{5}\sqcup\mbox{Free Hypers}
\]
Clearly, the $\mathbf{27}$ of $\mathfrak{e}_{6}$ has acquired a vev in this
case, and the $\mathfrak{e}_{6}$ symmetry has been spontaneously broken to
$\mathfrak{f}_{4}$.

This can be generalized to theories with multiple tensor multiplets. Here,
some combination of hypermultiplets may simultaneously be given vevs, so
multiple gauge symmetries may be broken at once. For instance, we may consider
the flow,
\[
\underset{[N_{f}=2]}{\overset{\mathfrak{su}_{3}}{2}} \,\, \underset{[N_{f}%
=5]}{\overset{\mathfrak{su}_{4}}{2}} \overset{RG}{\rightarrow}\underset{[N_{f}%
=1]}{\overset{\mathfrak{su}_{2}}{2}} \,\, \underset{[N_{f}%
=4]}{\overset{\mathfrak{su}_{3}}{2}} \sqcup\mbox{Free Hypers}
\]
Here, both gauge symmetries have been broken to subgroups by giving vevs to
charged hypermultiplets.

Computationally, it is clear from this that the the number of Higgs branch
flows will be enormous for a theory with many tensor multiplets and gauge
symmetries that are far from minimal. For tensor branch flows, we can simplify
the situation by decomposing any flow as the composition of several flows
which each involve only a single tensor multiplet. For instance, the flow
\[
1,2,2,2,2 \overset{RG}{\rightarrow}1,2 \sqcup2 \sqcup
\mbox{Two Free $(2,0)$ Tensors}
\]
may be decomposed as
\[
1,2,2,2,2 \overset{RG}{\rightarrow}1,2,2,2 \sqcup\mbox{Free $(2,0)$ Tensor}
\overset{RG}{\rightarrow}1,2 \sqcup2\sqcup\mbox{Two Free $(2,0)$ Tensors}
\]
by taking to infinity the vevs of the fifth and third tensor multiplet
scalars, respectively. Thus, in the case of tensor branch flows, it is enough
to consider only the flows initiated by giving a vev to a single tensor
multiplet. If some anomaly polynomial coefficient is positive under each of
these flows, it is guaranteed to be positive under their compositions, and
hence it will be positive under all flows.

However, the analogous statement is not true for Higgs branch flows. For
instance, the flow
\[
\lbrack N_{f}=3]\,\,{\overset{\mathfrak{su}_{3}}{2}}%
\,\,{\overset{\mathfrak{su}_{3}}{2}}\,\,{\overset{\mathfrak{su}_{3}}{2}%
}\,\,{\overset{\mathfrak{su}_{3}}{2}}\,\,[N_{f}=3]\overset{RG}{\rightarrow
}[N_{f}=2]\,\,{\overset{\mathfrak{su}_{2}}{2}}\,\,{\overset{\mathfrak{su}%
_{2}}{2}}\,\,{\overset{\mathfrak{su}_{2}}{2}}\,\,{\overset{\mathfrak{su}%
_{2}}{2}}\,\,[N_{f}=2]
\]
cannot proceed in a piecewise manner in which each $\mathfrak{su}_{3}$ is
Higgsed to $\mathfrak{su}_{2}$ one at a time. There is no way three of the
four nodes can hold an $\mathfrak{su}_{2}$ gauge algebra while the fourth
holds an $\mathfrak{su}_{3}$ gauge algebra without violating the convexity
condition of \cite{Heckman:2015bfa}, and so the last step in the process can
never be completed. Thus, not all Higgs branch flows can be decomposed as
single flows.

However, there is nothing stopping us from formally writing down the anomaly
polynomial for the would-be theory,
\[
[N_{f}=2] \,\, {\overset{\mathfrak{su}_{2}}{2}} \,\, \underset{[N_{f}%
=-1]}{\overset{\mathfrak{su}_{2}}{2}}\,\, \underset{[N_{f}%
=2]}{\overset{\mathfrak{su}_{2}}{3}}\,\, {\overset{\mathfrak{su}_{2}}{2}} \,\,
[N_{f}=1]
\]
The presence of $-1$ hypermultiplets on the second tensor multiplet indicates
that this theory does not exist, but nonetheless we may formally compute the
anomaly polynomial by simply subtracting the contribution $I_{\text{hyper}}$
for a hypermultiplet charged under that gauge symmetry. Once we allow for such
non-existent theories, which we henceforth call ``bogus theories," we may
indeed decompose these Higgs branch flows into sequences of single-node flows.
In the previous example, we have (neglecting the free, decoupled matter
sectors),
\[
[N_{f}=3]\,\, {\overset{\mathfrak{su}_{3}}{2}} \,\, {\overset{\mathfrak{su}%
_{3}}{2}} \,\, {\overset{\mathfrak{su}_{3}}{2}}\,\,{\overset{\mathfrak{su}%
_{3}}{2}} \,\, [N_{f}=3] \overset{RG}{\rightarrow}[N_{f}=1] \,\,
{\overset{\mathfrak{su}_{2}}{2}} \,\, \underset{[N_{f}%
=1]}{\overset{\mathfrak{su}_{3}}{2}}\,\, {\overset{\mathfrak{su}_{3}}{2}%
}\,\,{\overset{\mathfrak{su}_{3}}{2}} \,\, [N_{f}=3]
\]
\[
\overset{RG}{\rightarrow}[N_{f}=1] \,\, {\overset{\mathfrak{su}_{2}}{2}} \,\,
\underset{[N_{f}=1]}{\overset{\mathfrak{su}_{3}}{2}}\,\, \underset{[N_{f}%
=1]}{\overset{\mathfrak{su}_{3}}{2}}\,\,{\overset{\mathfrak{su}_{2}}{2}} \,\,
[N_{f}=1] \overset{RG}{\rightarrow}[N_{f}=2] \,\, {\overset{\mathfrak{su}%
_{2}}{2}} \,\, \underset{[N_{f}=-1]}{\overset{\mathfrak{su}_{2}}{2}}\,\,
\underset{[N_{f}=2]}{\overset{\mathfrak{su}_{3}}{2}}%
\,\,{\overset{\mathfrak{su}_{2}}{2}} \,\, [N_{f}=1]
\]
\[
\overset{RG}{\rightarrow}[N_{f}=2] \,\, {\overset{\mathfrak{su}_{2}}{2}} \,\,
{\overset{\mathfrak{su}_{2}}{2}}\,\, {\overset{\mathfrak{su}_{2}}{2}%
}\,\,{\overset{\mathfrak{su}_{2}}{2}} \,\, [N_{f}=2]
\]
Of course, there are multiple sequences with the same endpoints, but the
difference of the anomaly polynomial coefficients at the beginning and end of
the flow depend only on the endpoints and so are independent of the path
taken. As long as some polynomial coefficient is monotonically decreasing at
each step of the sequence, then it is guaranteed to decrease along the entire
flow. This allows us to perform a large, systematic sweep of Higgs branch flows.

The Higgs branch flows discussed so far all preserve the structure of the
tensor branch i.e. the F-theory base. However, there are other, more
complicated flows which do not \cite{Heckman:2015ola}. For
example, separating two $M5$-branes probing an $E_{8}$-wall along a direction
parallel to the wall corresponds to the flow,
\[
1,2\rightarrow1\sqcup1\sqcup\mbox{Free $(2,0)$ Tensor}
\]
This is a Higgs branch flow, but it clearly does not preserve the tensor
multiplet structure.

\subsection{Flows for the Sweeps}

Although the form of 6D SCFTs is highly constrained, there are still too many
theories to perform a fully systematic sweep of RG flows. Our plan will be to
focus on SCFTs with a single tensor node, as well as
the classical theories of reference \cite{Heckman:2015bfa} (see
also \cite{Bhardwaj:2015xxa}). Since we will be restricting our attention to these cases,
let us briefly explain why we expect these theories in particular to give us
the tightest bounds.

In the case of a generalized quiver, we will typically need to discuss moving
onto the tensor branch or Higgs branch for a system by eliminating some amount
of conformal matter. Now, the important feature of conformal matter is that it
typically is built of smaller SCFTs which themselves contain a large number of
degrees of freedom. From this perspective, we expect to obtain the sharpest
bounds from theories where the drop in the number of degrees of freedom is the
smallest. These are cases where the matter fields are genuine weakly coupled
hypermultiplets (in the tensor branch description).

Instead, we concentrate on tensor branch flows between (a) classical theories
and (b) UV theories of a single tensor node to IR theories without a tensor
node, and we concentrate on Higgs branch flows (which preserve the tensor
multiplet structure) between (c) classical theories and (d) theories of a
single tensor node. From these, we map out the monotonic region of m-space. We
will find that this region is rather sizeable, indicating a large family of
monotonically decreasing functions.

``Classical theories" are defined to be theories built in
F-theory from curves whose self-intersection is either $-1$, $-2$, or $-4$,
contain no spinors, and have no exceptional fiber types (e.g. $II$, $II$,
$IV$, etc.). The classical theories are easily classified and come in the
following families:
\[
\overset{\mathfrak{su}({n_{1}})}{2}\,\,...\,\,\overset{\mathfrak{su}({n_{k}%
})}{2}%
\]%
\[
\overset{\mathfrak{su}({n_{1}})}{2} \,\, \underset{\mathfrak{su}({n_{t}})}{\underset{2}{\overset{\mathfrak{su}({n_{2}})}{2}}} \,\, ...\,\,\overset{\mathfrak{su}({n_{k}%
})}{2}%
\]%
\[
\overset{\mathfrak{su}({n_{1}})}{2}\overset{\mathfrak{su}({n_{2}})}{2} \,\, \underset{\mathfrak{su}({n_{t}})}{\underset{2}{\overset{\mathfrak{su}({n_{3}})}{2}}} \,\, ...\,\,\overset{\mathfrak{su}({n_{k}%
})}{2}%
\]%
\[
\overset{\mathfrak{su}({n_{0}})}{1}\,\,\overset{\mathfrak{su}({n_{1}}%
)}{2}\,\,...\,\,\overset{\mathfrak{su}({n_{k}})}{2}%
\]%
\[
\overset{\mathfrak{sp}({n_{0}})}{1}\,\,\overset{\mathfrak{su}({n_{1}}%
)}{2}\,\,...\,\,\overset{\mathfrak{su}({n_{k}})}{2}%
\]%
\[
\overset{\mathfrak{su}({m})}{2}\,\,\overset{\mathfrak{sp}({n_{0}}%
)}{1}\,\,\overset{\mathfrak{so}({n_{1}})}{4}\,\,\overset{\mathfrak{sp}({n_{2}%
})}{1}\,\,...\,\,\overset{\mathfrak{sp}({n_{k-1}})}{1}%
\,\,\overset{\mathfrak{so}({n_{k}})}{4}%
\]%
\[
\overset{\mathfrak{su}({m})}{2}\,\,\overset{\mathfrak{sp}({n_{0}}%
)}{1}\,\,\overset{\mathfrak{so}({n_{1}})}{4}\,\,\overset{\mathfrak{sp}({n_{2}%
})}{1}\,\,...\,\,\overset{\mathfrak{so}({n_{k-1}})}{4}%
\,\,\overset{\mathfrak{sp}({n_{k}})}{1}%
\]%
\[
\overset{\mathfrak{sp}({n_{0}})}{1}\,\,\overset{\mathfrak{so}({n_{1}}%
)}{4}\,\,\overset{\mathfrak{sp}({n_{2}})}{1}\,\,...\,\,\overset{\mathfrak{so}%
({n_{k-1}})}{4}\,\,\overset{\mathfrak{sp}({n_{k}})}{1}%
\]%
\[
\overset{\mathfrak{sp}({n_{0}})}{1}\,\,\overset{\mathfrak{so}({n_{1}}%
)}{4}\,\,\overset{\mathfrak{sp}({n_{2}})}{1}\,\,...\,\,\overset{\mathfrak{sp}%
({n_{k-1}})}{1}\,\,\overset{\mathfrak{so}({n_{k}})}{4}%
\]%
\[
\overset{\mathfrak{so}({n_{1}})}{4}\,\,\overset{\mathfrak{sp}({n_{2}}%
)}{1}\,\,...\,\,\overset{\mathfrak{sp}({n_{k-1}})}{1}%
\,\,\overset{\mathfrak{so}({n_{k}})}{4}%
\]%
\[
\overset{\mathfrak{sp}({n_{0}})}{1}\,\,\underset{\mathfrak{sp}%
(m)}{\underset{1}{\overset{\mathfrak{so}({n_{1}})}{4}}}%
\,\,\overset{\mathfrak{sp}({n_{2}})}{1}\,\,...\,\,\overset{\mathfrak{so}%
({n_{k-1}})}{4}\,\,\overset{\mathfrak{sp}({n_{k}})}{1}%
\]%
\[
\overset{\mathfrak{sp}({n_{0}})}{1}\,\,\underset{\mathfrak{sp}%
(m)}{\underset{1}{\overset{\mathfrak{so}({n_{1}})}{4}}}%
\,\,\overset{\mathfrak{sp}({n_{2}})}{1}\,\,...\,\,\overset{\mathfrak{sp}%
({n_{k-1}})}{1}\,\,\overset{\mathfrak{so}({n_{k}})}{4}%
\]%
\[
\overset{\mathfrak{so}({n_{1}})}{4}\,\,\overset{\mathfrak{sp}({n_{2}}%
)}{1}\,\,\underset{\mathfrak{sp}(m)}{\underset{1}{\overset{\mathfrak{so}%
({n_{3}})}{4}}}\,\,\overset{\mathfrak{sp}({n_{4}})}{1}%
\,\,\overset{\mathfrak{so}({n_{5}})}{4}%
\]
The gauge algebras in all of these theories obey the convexity conditions
discussed in \cite{Heckman:2015bfa}, which we will not repeat here.

We initiate a tensor branch flow by taking the size of one or more of the curves to infinity. It suffices to
consider the case in which only a single curve is taken large, as a general
flow can be expressed as a composition of such flows. Thus, for each of the
families of flows considered, we list out all theories up to a given number of
tensor nodes. For each theory,
we consider the flow induced by taking each node large in turn.

Similarly, a Higgs branch flow occurs when we decrease the rank of the gauge
algebras living on one or more of the tensor nodes. As discussed previously,
decreasing the rank of the gauge algebra of just a single tensor node will
sometimes take us outside the class 6D SCFTs, but we may compensate for this
by formally defining anomaly polynomials for \textquotedblleft bogus
theories." Thus, for each family of classical theories, we start with all
theories (real or bogus) up to a given number of tensor nodes. For each theory, we consider the flow
induced by minimally decreasing the gauge algebra on each node in turn and use numerical optimization to determine the tightest possible bound such a flow could give for any possible gauge algebra rank.

\section{Analysis and Results \label{sec:ANALYSIS}}

We now turn to a summary of our automated sweeps.
We find that the strongest constraints on the change in the conformal
anomalies come from the \textit{simplest} theories. This means that our brute
force sweeps actually amount to strong evidence in favor of the
existence of such C-theorems.

As already mentioned, we focus on linear combinations:
\begin{equation}
C=\overrightarrow{m}\cdot\overrightarrow{\alpha},
\end{equation}
where $\overrightarrow{\alpha}=(\alpha,\beta,\gamma,\delta)$
is the vector of anomaly polynomial coefficients. Our goal is to study the
region in the 4-dimensional space of $(m_{1},m_{2},m_{3},m_{4})$ such that $C$
is monotonically decreasing under RG flows. We will refer to this region as
the \textquotedblleft monotonic region" of m-space. At times, we will
distinguish the region that is monotonic under tensor branch flows from the
region that is monotonic under Higgs branch flows. The region that is
monotonic under both types of flows is then given by the intersection of these
two monotonic regions.

From the form of the anomaly polynomial differences under tensor branch and
Higgs branch flows considered earlier, it is clear that the the $m_{4}$
direction will be a null direction, since $\Delta\delta= 0$ under all such
flows. This reduces the parameter space of interest to the three-dimensional
space of $(m_{1}, m_{2}, m_{3},0)$. In the case of Higgs branch flows, the
$m_{3}$ direction will also be null, but for tensor branch flows $\Delta
\gamma\neq0$.

Our analysis consists of a large sweep over the classical theories. For Higgs branch flows, we scan over each classical configuration consisting of 25 or fewer tensor nodes.  For each such configuration, we consider the bound placed on the monotonic region by minimally lowering one of the gauge algebra ranks.  We then numerically minimize the quantity $-\Delta\alpha/\Delta\beta$ over the gauge algebra ranks for each possible flow.\footnote{All numerical optimizations described in this paper were performed using \texttt{Mathematica}.}  The bound on the monotonic region from any Higgs branch flow is given by $m_2 / m_1 < -\Delta\alpha/\Delta\beta$, so by minimizing this value, we determine the tightest possible bound on the monotonic region.

For example, the flow
\[
\underset{[N_{f}=2N_L - N_R]}{\overset{\mathfrak{su}_{N_L}}{2}} \,\, \underset{[N_{f}%
=2 N_R - N_L]}{\overset{\mathfrak{su}_{N_R}}{2}} \overset{RG}{\rightarrow} \underset{[N_{f}=2N_L - N_R-2]}{\overset{\mathfrak{su}_{N_L-1}}{2}} \,\, \underset{[N_{f}%
=2 N_R - N_L+1]}{\overset{\mathfrak{su}_{N_R}}{2}} \sqcup\mbox{Free Hypers}
\]
yields an anomaly polynomial difference vector of $(\Delta\alpha,\Delta\beta,\Delta\gamma,\Delta\delta) = (\frac{7 N_L}{12}+\frac{N_R}{3}- \frac{7}{24},\frac{1}{48}-\frac{N_L}{24},0,0)$.  For such a flow, we must have $N_L \geq 2$, $N_R \geq 1$, so we minimize the quantity $-\Delta\alpha/\Delta\beta = 14 + \frac{16 N_R}{ 2 N_L-1}$ subject to these constraints.  The minimum value of this quantity is $14$ in this example, and hence we find a bound $m_2/m_1 < 14$.

The theories consisting of a single tensor node are listed in section 6.1 of \cite{Heckman:2015bfa}, and we will not repeat them here.  The single node Higgs branch flows begin with a UV theory of a single tensor node and flow to an IR theory of a single tensor node via Higgsing of the gauge algebra.  The tensor branch structure is unaffected.  Our sweep encompasses all such flows as well as the flow of the E-string theory to free hypermultiplets, $1 \overset{RG}{\rightarrow}$ Free Hypers.  This last flow lies at the bottom of the hierarchy of Higgs branch flows involving theories with single tensor multiplets and ends up providing the only meaningful bound of any Higgs branch flow.

For tensor branch flows, we similarly scan over every classical configuration consisting of 25 or fewer tensor nodes.  For each such configuration, we consider the bound placed on the monotonic region by taking the vev of one tensor multiplet scalar to infinity.  We then numerically optimize over the gauge algebra ranks to determine the tightest possible bound on the monotonic region.

Whereas in the case of Higgs branch flows it sufficed to minimize the quantity $-\Delta\alpha/\Delta\beta$ along all flows and show that it increases with the number of tensor nodes, here there are three quantities which must be considered.  Our goal is to study the bounds in the range $-116/9<m_2/m_1 < 26/11$, and we want to know if any flows will provide tighter bounds than the ones in (\ref{tensor1}) and (\ref{tensor2}).  To ensure that all such bounds are trivial in this range, it suffices to check that each of the following hold:
\begin{subequations}
\begin{align}
-\frac{\Delta\alpha}{\Delta\beta}>2 \text{ OR }  -\frac{\Delta\alpha}{\Delta\beta}<-6 \label{tensorcheck1}\\
b := -\frac{26}{11} \frac{\Delta\beta}{\Delta\gamma} - \frac{\Delta\alpha}{\Delta\gamma} < \frac{32}{11} \label{tensorcheck2}\\
c : =\frac{116}{9} \frac{\Delta\beta}{\Delta\gamma} - \frac{\Delta\alpha}{\Delta\gamma} < \frac{496}{3}  \label{tensorcheck3}
\end{align}
\end{subequations}
These three conditions together ensure that none of the bounding lines $\Delta\alpha + \Delta\beta m_2/m_1 + \Delta\gamma m_3/m_1 =0$ coming from a tensor branch flow cross above either of the two lines shown in Figure \ref{Tensorplot}, meaning that these two flows do indeed provide the tightest bounds on the monotonic region.

We also consider tensor branch flows for theories consisting of a single tensor node.  These flows are very simple: one takes the vev of the tensor multiplet scalar to infinity and is left with a free tensor multiplet and an appropriate number of free vectors and free hypers in the IR.  Once again, our sweep encompasses all such flows.

One may worry that we are overestimating the size of the monotonic region due to the limited scope of our analysis.  However, we will soon see that the strongest constraints on monotonicity come from the simplest flows, so it appears highly unlikely to us that the more complicated examples we have abstained from considering will provide meaningful bounds.

In fact, our analysis is more likely to \emph{under}estimate the size of the monotonic region due to the fact that we are including ``bogus theories" in our analysis.  Any flow beginning or ending on a bogus theory is a ``bogus flow," which does not actually exist yet may introduce a constraint on the monotonic region.  Any function which is monotonic under all (possibly bogus) flows will necessarily be monotonic under all real flows, so the true monotonic region may be underestimated using these methods.

However, it turns out that the tighest constraints from our sweep come from honest tensor branch flows and Higgs branch flows, which do not involve any bogus theories, so it appears that we are neither underestimating nor overestimating the size of the monotonic region.

We now examine the monotonic region for both Higgs branch and tensor branch flows.

\subsection{Higgs Branch Flows}

\begin{figure}
\begin{center}
\includegraphics[trim=0mm 0mm 0mm 0mm, clip, width=140mm]{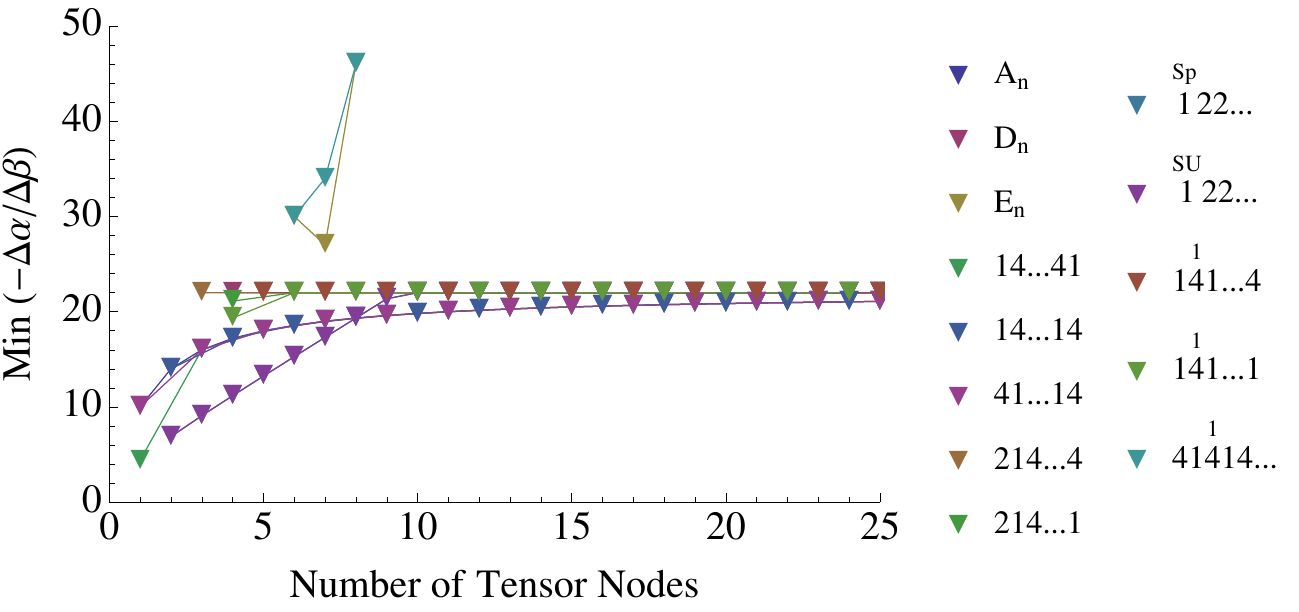}
\caption{The minimum value of $-\Delta\alpha/\Delta\beta$ for each classical family, up to 25 tensor nodes.  Since each value of this quantity provides a bound $m_2 / m_1 < -\Delta\alpha/\Delta\beta$, only the smallest value is meaningful, while the rest of the bounds are redundant.  Clearly, the bounds weaken as the number of tensor branch nodes increases, giving us strong reason to believe that the only meaningful bound comes from (\ref{Higgsbound}).}
\label{Higgstrend}
\end{center}
\end{figure}

The two key observations from our numerical sweep of Higgs branch flows are as follows:

\begin{enumerate}
\item The constraints on the monotonic region given by a Higgs branch flow from one theory to a nearby theory grow weaker as the complexity of the theories increases.
\item The monotonic region for Higgs branch flows comes from a single flow. It is given by moving onto the Higgs branch of the rank one E-string theory. On the Higgs branch, we are left with $30$ free hypermultiplets (including the center of mass degrees of freedom), as dictated by matching the coefficient $p_2(T)$ of the anomaly polynomial. This yields the bound:\footnote{Here and henceforth, we are assuming $m_1 >0$ in the monotonic region.  We will see that this condition is necessarily enforced by tensor branch flows, so it is safe to assume.}
\begin{align}
1 \overset{RG}\rightarrow\mbox{ Free Hypers.} \Rightarrow 0 &< m_1 - \frac{11 m_2}{26}  \label{Higgsbound}
\end{align}
\end{enumerate}
The first of these observations is crucial, for it gives us strong reason to believe that the monotonic region will not shrink once more complicated Higgs branch flows are considered.  As the gauge algebra rank is increased, the self-intersection number of the curve is decreased, or the number of tensor nodes is increased, the constraints on the monotonic region are strictly weaker.  This is illustrated in Figure \ref{Higgstrend}, which shows how the quantity $-\Delta\alpha/\Delta\beta$ and hence the bound $m_2 / m_1 < -\Delta\alpha/\Delta\beta$ varies with the number of tensor nodes for each of the classical families considered.  Clearly, the bounds grow monotonically weaker with the number of tensor nodes, eventually leveling off.  This gives us strong grounds to believe that (\ref{Higgsbound}) is the only constraint on the monotonic region for Higgs branch flows.

\begin{figure}
\begin{center}
\includegraphics[trim=0mm 0mm 0mm 0mm, clip, width=100mm]{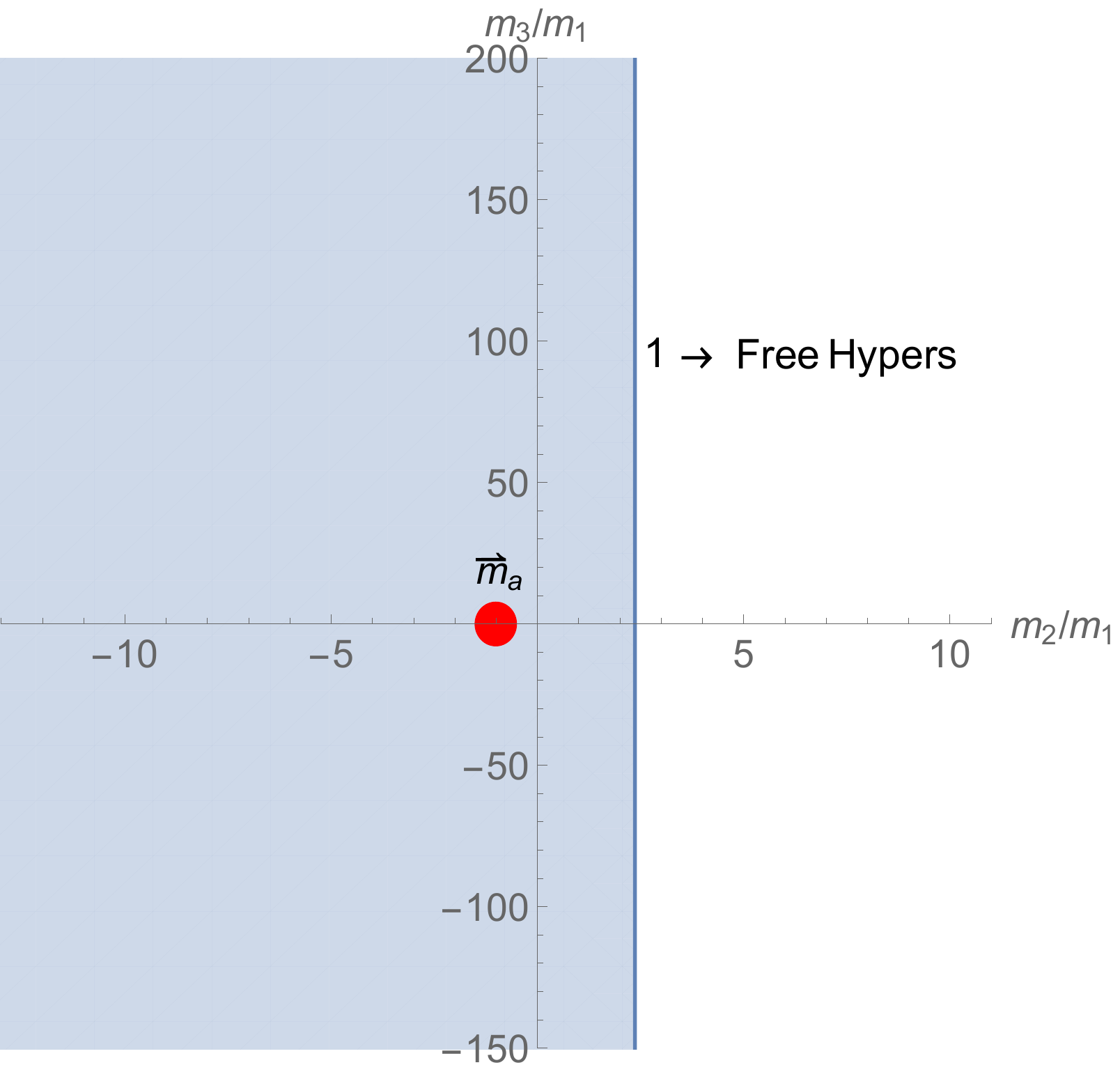}
\caption{The bounds on the monotonic region $(m_1,m_2,m_3,m_4=0)$ from Higgs branch flows. The only meaningful bound is $m2 / m_1 < 26/11$, and it comes from the simplest possible Higgs branch flow, $1 \overset{RG}\rightarrow \mbox{ Free Hypers.}$}
\label{Higgsfigure}
\end{center}
\end{figure}

Assuming this is accurate, it would establish the a-theorem for Higgs branch flows.  In \cite{Cordova:2015fha} (see also \cite{EHOLOG}), it was shown that the $a$-type Weyl anomaly corresponds to the vector,
\begin{equation}
\overrightarrow{m}_a =\frac{8}{3}(1,-1,1,\frac{3}{8})
\end{equation}
Thus, $m_2 / m_1 = -1 < 26/11$, so $\overrightarrow{m}_a$ lies comfortably within the monotonic region for Higgs branch flows.  This is shown in Figure \ref{Higgsfigure}.

\subsection{Tensor Branch Flows}

We now present the two key observations from our analysis of tensor branch flows.  The first of these is an analytic result, whereas the second is an observation from our numerical sweep:

\begin{enumerate}
\item The monotonic region for tensor branch flows has $m_1 >0$ and is well approximated by the inequality $m_1 m_3 > m_2^2$.
\item Within the range $-116/9 < m_2/m_1 < 26/11$, the bounds appear to grow weaker as the complexity of the theories involved increases, and the most stringent bounds from our numerical sweep are
\begin{subequations}
\label{monotonicbounds}
\begin{align}
1 \overset{RG}\rightarrow \mbox{Free Tensor} \Rightarrow 0 <& m_1 - \frac{m_2}{2} + \frac{m_3}{16} \label{bound1} \\
\overset{\mathfrak{e}_8}{12} \mbox{ or } \overset{\mathfrak{e}_7}{8} \mbox{ or } \overset{\mathfrak{e}_6}{6} \mbox{ or } \overset{\mathfrak{f}_4}{5} \mbox{ or } \overset{\mathfrak{so}_8}{4} \mbox{ or } \overset{\mathfrak{su}_3}{3} \overset{RG}\rightarrow \mbox{Free Tensor } \decoup & \mbox{ Free Vectors} \Rightarrow 0 < m_1 + \frac{m_2}{6} + \frac{m_3}{144}
\label{bound2} \\
2 \overset{RG}\rightarrow \mbox{ Free Tensor} \decoup  \mbox{ Free } &\mbox{Hypers} \Rightarrow 0 < m_1 \label{bound3}
\end{align}
\end{subequations}
\end{enumerate}
The flow in (\ref{bound3}) has $\Delta\beta = \Delta\gamma =0$, $\Delta\alpha > 0$ and hence justifies the bound $m_1 > 0$ on the monotonic region.

\begin{figure}
\begin{center}
\includegraphics[trim=0mm 0mm 0mm 0mm, clip, width=120mm]{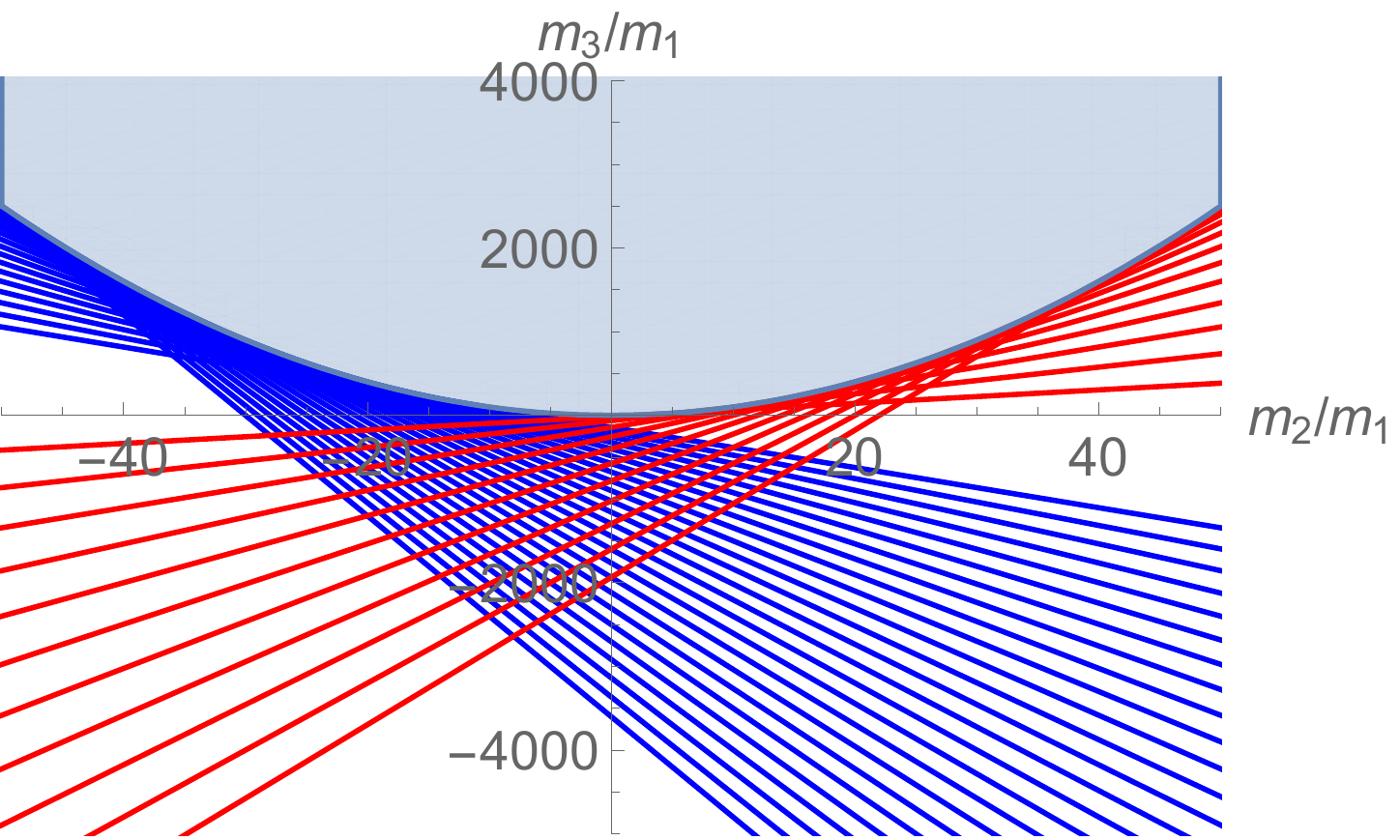}
\caption{The bounds on the monotonic region $(m_1,m_2,m_3,m_4=0)$ from tensor branch flows of $\underset{[N_f = n - 8]}{\overset{\mathfrak{so}(n)}4}$ theories (blue, negative slope lines) and $\underset{[N_f = 8n + 2]}{\overset{\mathfrak{sp}(n)}1}$ theories (red, positive slope lines).  For large $|m_2 / m_1|$, the shaded region $m_1 m_3 > m_2^2$ is well approximated by the bounds from these flows.}
\label{tangent}
\end{center}
\end{figure}

Recall in section \ref{sec:FLOWS} we argued for the analytic bound $m_2^2 < m_1 m_3$. Thus,
the monotonic region for tensor branch flows is necessarily a subset of this region.
In fact, this region provides a rather good approximation to the monotonic region for tensor branch flows.  The flow
$$
\overset{\mathfrak{so}_n}{4} \overset{RG}\rightarrow \mbox{Free Tensor} \decoup \mbox{ Free Vectors}  \decoup \mbox{ Free Hypers}
$$
gives the bound
\begin{equation}
0 < 4 (n-2)^2 m_1 + 4 (n-2) m_2 + m_3.
\label{4tensorflows}
\end{equation}
Similarly, the flow
$$
\underset{N_f = 10}{\overset{\mathfrak{sp}_n}{1}} \overset{RG}\rightarrow \mbox{Free Tensor} \decoup \mbox{ Free Vectors}  \decoup \mbox{ Free Hypers}
$$
gives the bound
\begin{equation}
0 < 16 (n+1)^2 m_1 - 8 (n+1) m_2 + m_3.
\label{1tensorflows}
\end{equation}
Since $m_1>0$, we can without loss of generality set $m_1 = 1$, after which these bounds correspond to lines in the $m_2$-$m_3$ plane.  This is shown in Figure \ref{tangent}.  Clearly, each of these lines is tangent to the parabola $m_3 = m_2^2$, which upon restoring $m_1$ is just the hypersurface $m_1 m_3 = m_2^2$.  Recall that the monotonic region for tensor branch flows contains the region $m_1 m_3 > m_2^2$ bounded by this hypersurface.

If we were to analytically continue the formulae (\ref{4tensorflows}) and (\ref{1tensorflows}) to arbitrary rational $n$, we would find that the region they bound is precisely $m_1 m_3> m_2^2$.  However, in reality, $\mathfrak{so}_n$ on a $-4$ curve makes sense only for integral $n \geq 8$ and $\mathfrak{sp}_n$ on a $-1$ curve makes sense only for integral $n \geq 0$.  Thus, the monotonic region for tensor branch flows is actually slightly larger than the region $m_1 m_3> m_2^2$, though as Figure \ref{tangent} illustrates, it is still a good approximation.  Furthermore, our sweeps reveal that tensor branch flows for theories with more than one tensor node or non-classical configurations sometimes provide non-trivial bounds.  In other words, the monotonic region is even slightly tighter than the region given by (\ref{4tensorflows}) and (\ref{1tensorflows}).

\begin{figure}
\begin{center}
\includegraphics[trim=0mm 0mm 0mm 0mm, clip, width=120mm]{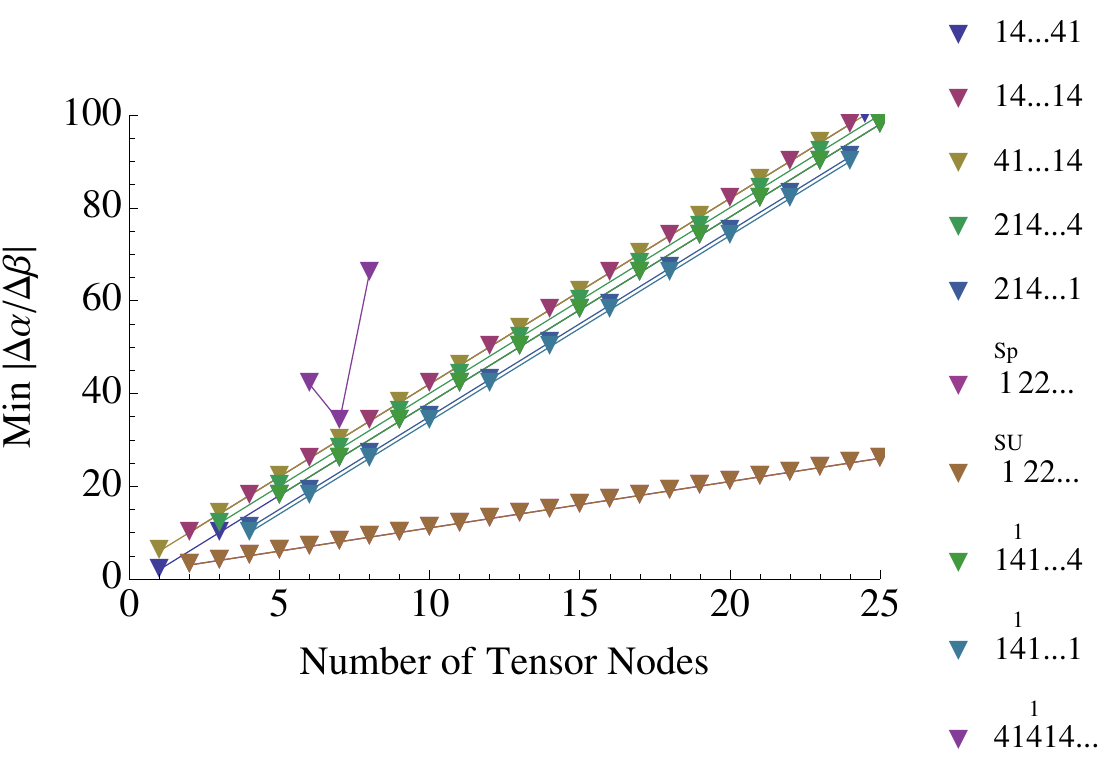}
\caption{The minimum value of $|\Delta\alpha/\Delta\beta|$ for each classical family, up to 25 tensor nodes.  The values increase monotonically with the number of tensor nodes for each classical family, indicating that the bounds are getting weaker as the number of tensor nodes increases.}
\label{Tensortrend}
\end{center}
\end{figure}

\begin{figure}
\begin{center}
\includegraphics[trim=0mm 0mm 0mm 0mm, clip, width=120mm]{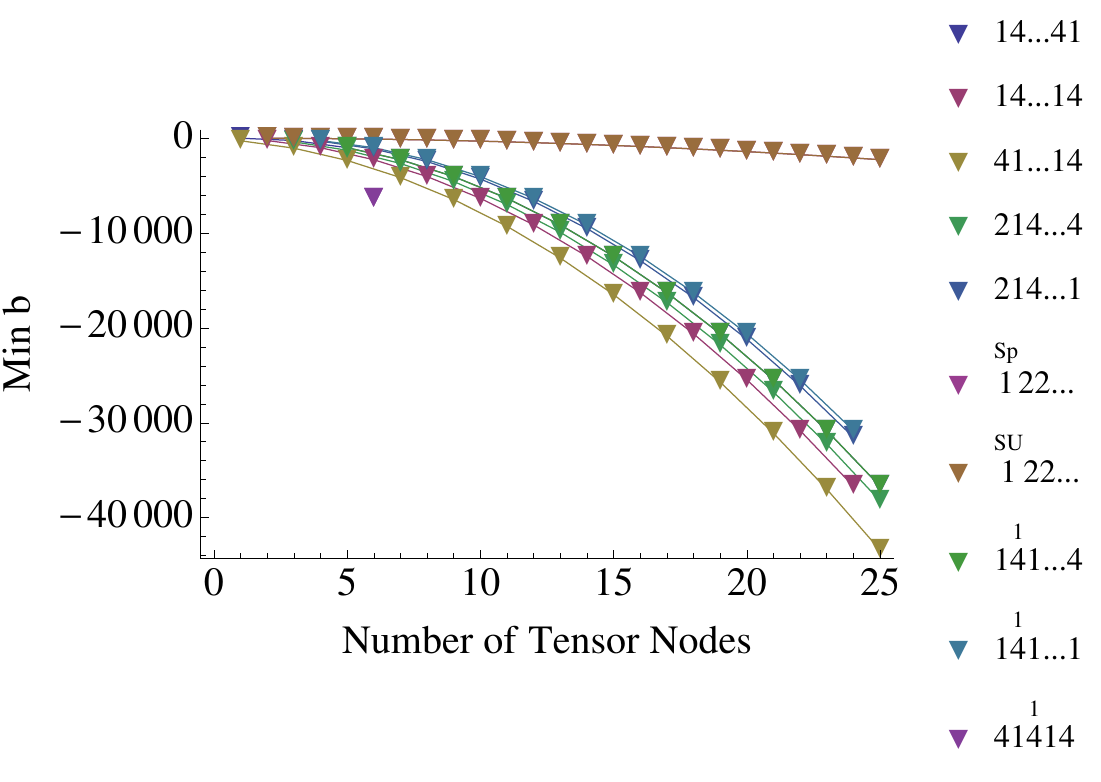}
\caption{The maximum value of $b$ for each classical family, up to 25 tensor nodes.  The values decrease monotonically with the number of tensor nodes for each classical family, indicating that the bounds are getting weaker as the number of tensor nodes increases.}
\label{Tensortrendb}
\end{center}
\end{figure}

\begin{figure}
\begin{center}
\includegraphics[trim=0mm 0mm 0mm 0mm, clip, width=120mm]{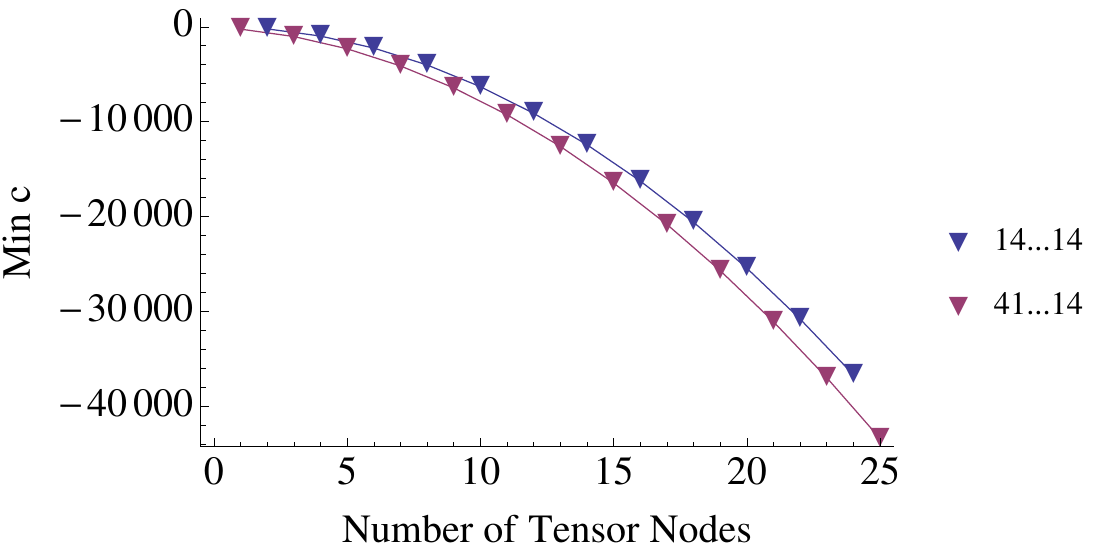}
\caption{The maximum value of $c$ for each classical family for which $\Delta\beta > 0$ under tensor branch flows, up to 25 tensor nodes.  The values decrease monotonically with the number of tensor nodes for each classical family, indicating that the bounds are getting weaker as the number of tensor nodes increases.}
\label{Tensortrendc}
\end{center}
\end{figure}

However, we also observe that the non-trivial bounds on $m_3$ offered by more complicated flows appear negligible for sufficiently small $| m_2/m_1|$.  If we focus on the region $-116/9 < m_2/m_1 < 26/11$,\footnote{Note that this number $26/11$ is chosen because $m_2 / m_1 < 26/11$ is the bound we observed on the monotonic region from Higgs branch flows.  The number $-116/9$ is chosen because a different tensor branch flow provides a stronger bound once $m_2 / m_1 < -116/9$.} we find that the only non-trivial bounds in our numerical sweep come from the simplest tensor branch flows depicted in (\ref{monotonicbounds}).  All other RG flows considered give strictly weaker bounds provided we concentrate on the region $-116/9 < m_2/m_1 < 26/11$.  In other words, starting with these incredibly simple flows, we find no new bounds upon increasing the number of tensor multiplets or the gauge group ranks.  In fact, the quantity $|\Delta\alpha/\Delta\beta|$ of (\ref{tensorcheck1}) is monotonically increasing with the number of tensor nodes across each classical family, while the quantities $b$ and $c$ of (\ref{tensorcheck2}) and (\ref{tensorcheck3}) are monotonically decreasing, so the bounds on the monotonic region grow monotonically weaker with the number of nodes in the region $-116/9 < m_2/m_1 < 26/11$.  This is shown in Figures \ref{Tensortrend}-\ref{Tensortrendc}.  Note that most of the families do not have any flows with $\Delta\beta > 0$, and these will trivially satisfy (\ref{tensorcheck3}), hence they are not shown here.  Also, all of the tensor branch flows with only $-2$ curves have $\Delta\alpha >0$, $\Delta\beta = \Delta\gamma = \Delta\delta = 0$, so these all provide the same bound $m_1 > 0$.  The families with only $-2$ curves are thus not included in any of the plots.

\begin{figure}
\begin{center}
\includegraphics[trim=0mm 0mm 0mm 0mm, clip, width=120mm]{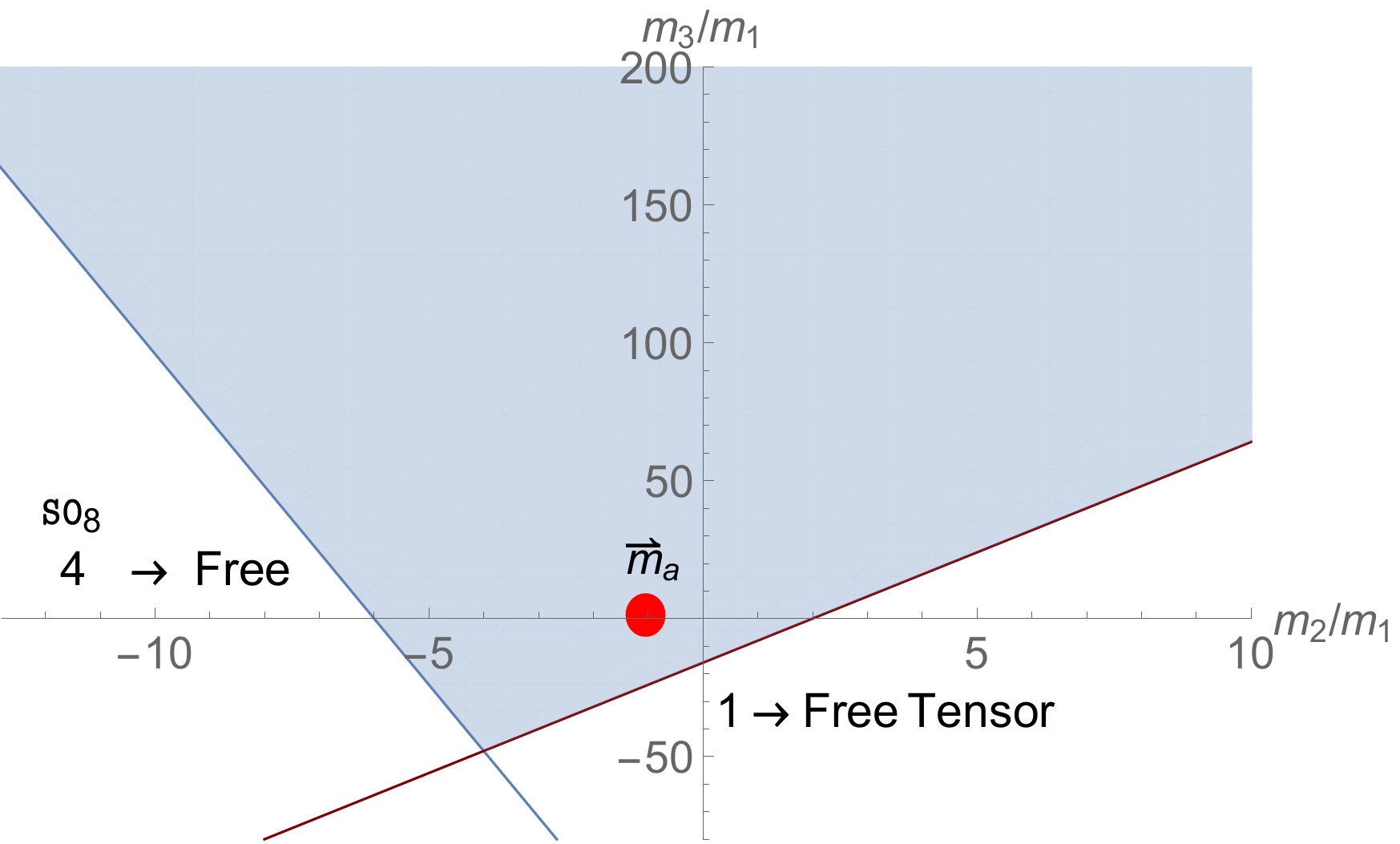}
\caption{The bounds on the monotonic region $(m_1,m_2,m_3,m_4=0)$ (shaded) from tensor branch flows.  The only meaningful restrictions come from the simple flows shown and give the bounds in (\ref{monotonicbounds}).   The vector $\vec{m_a}$ for the $a$-type Weyl anomaly (red) fits comfortably in the monotonic region.}
\label{Tensorplot}
\end{center}
\end{figure}

The resulting monotonic region is shown graphically in Figure \ref{Tensorplot}.  We have also indicated on this plot the location of $\overrightarrow{m}_a$, the vector for the $a$-type Weyl anomaly.  As expected, it lies well within the monotonic region for tensor branch flows as well as Higgs branch flows.

\subsection{Combined Analysis}

Finally, we combine the results of our Higgs branch and tensor branch analyses.  The resulting monotonic region is simply the intersection of the regions for each of these two types of flows.  We note that it fills a very large portion of m-space and contains $\overrightarrow{m}_a$.  We have plotted the resulting monotonic region in both the $|m_2/m_1|$ small and $|m_2/m_1|$ large limits.

\begin{figure}[H]
\begin{center}
\includegraphics[trim=0mm 0mm 0mm 0mm, clip, width=90mm]{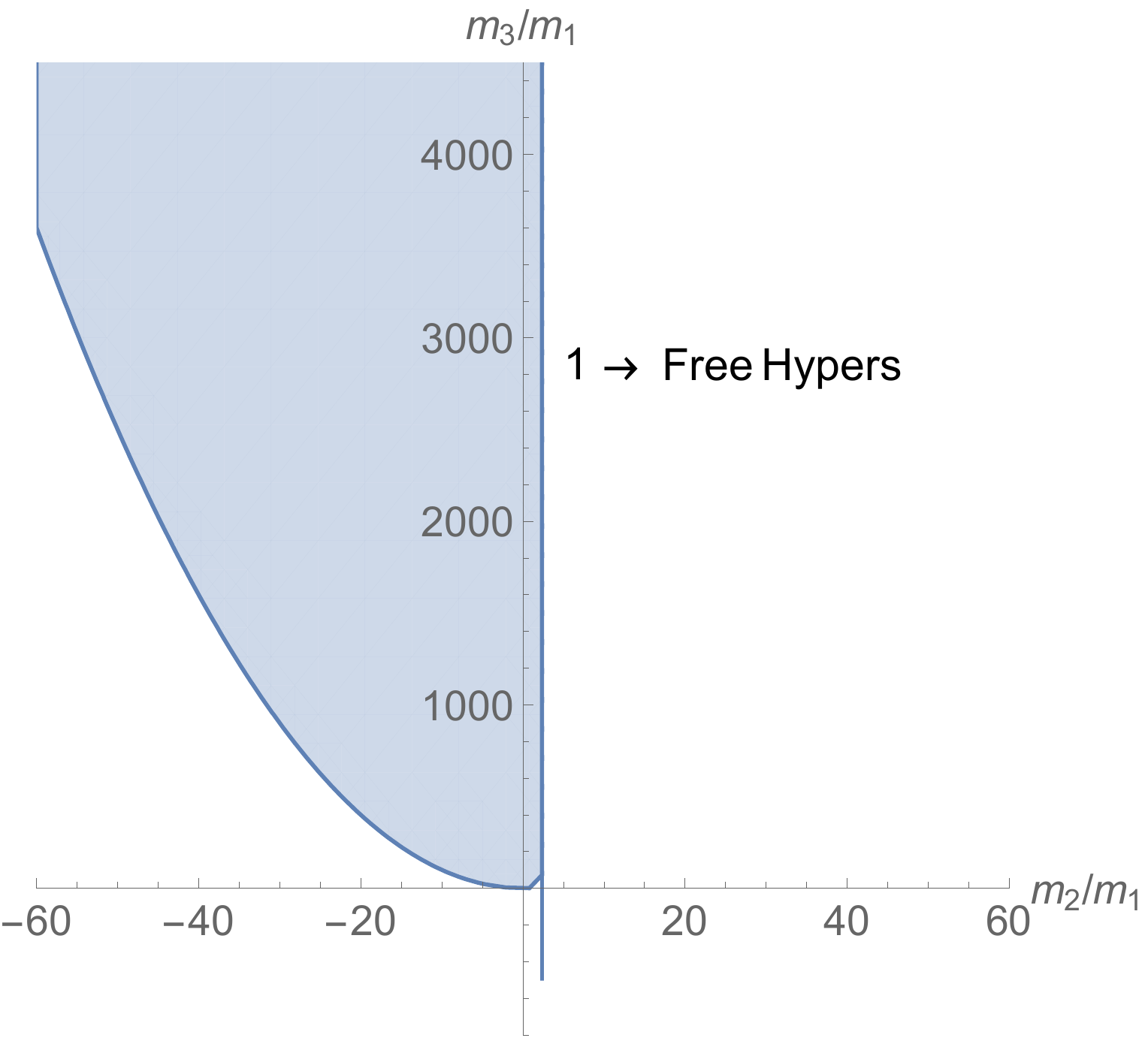}
\caption{The bounds on the monotonic region $(m_1,m_2,m_3,m_4=0)$ (shaded) from tensor and Higgs branch flows for large $|m_2 / m_1|$.  The region fills a large portion of m-space and is well approximated by the inequalities $m_1 m_3 > m_2^2$, $ 26/11 < m_2 / m_1 $.}
\label{weakerbounds}
\end{center}
\end{figure}

\begin{figure}[H]
\begin{center}
\includegraphics[trim=0mm 0mm 0mm 0mm, clip, width=120mm]{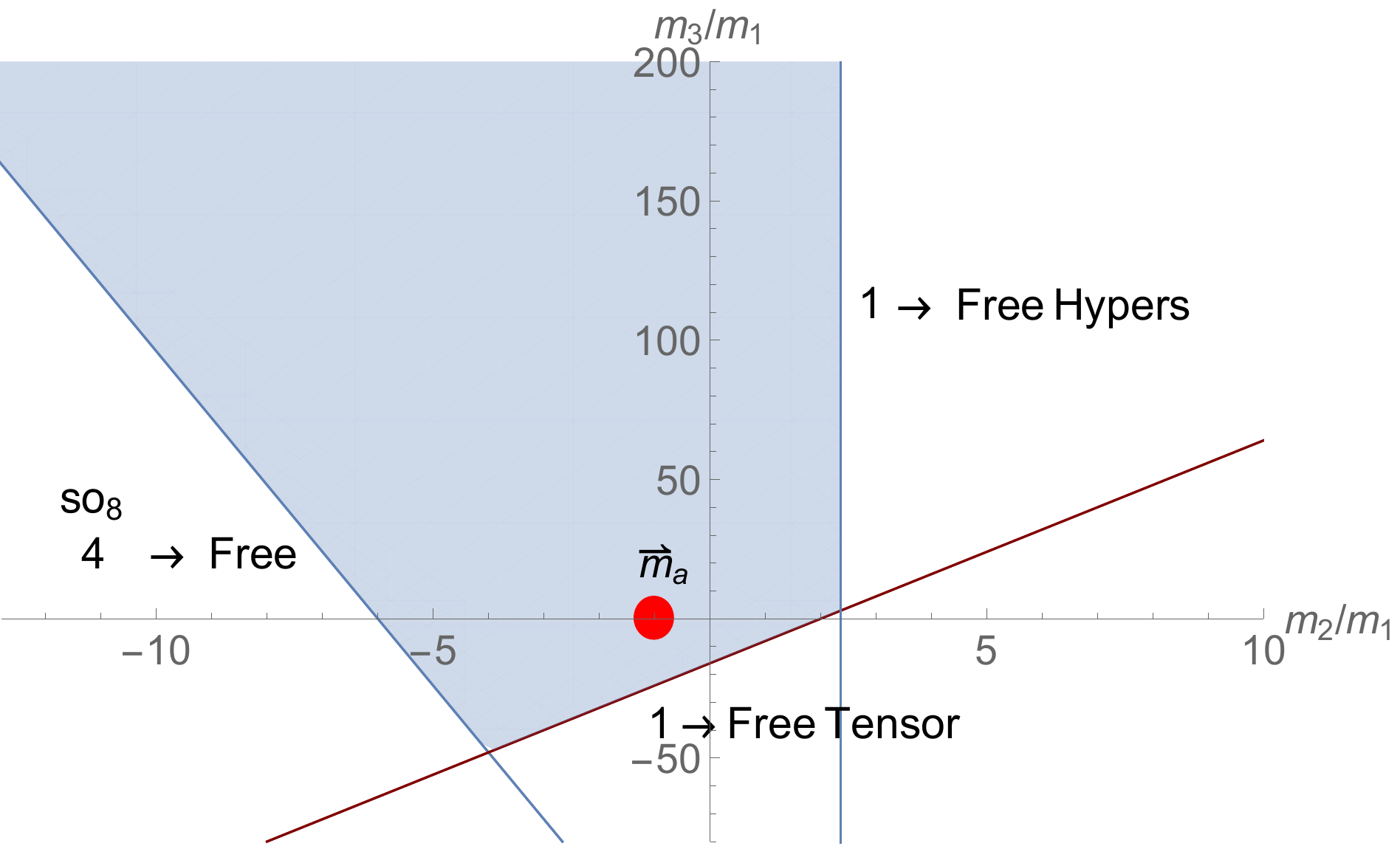}
\caption{The bounds on the monotonic region $(m_1,m_2,m_3,m_4=0)$ (shaded) from tensor and Higgs branch flows for $-116/9 < m_2/m_1 < 26/11$.  The only meaningful restrictions come from the simple flows shown and give the bounds in (\ref{Higgsbound}) and (\ref{monotonicbounds}).   The vector $\vec{m_a}$ for the $a$-type Weyl anomaly (red) fits comfortably in the monotonic region.}
\label{weakerbounds2}
\end{center}
\end{figure}

\section{Conclusions \label{sec:CONC}}

The existence of a C-function for CFTs provides a way to quantify the loss of
degrees of freedom in flows from the UV to the IR. In this note we have
considered candidate C-functions constructed from linear combinations of the
anomaly polynomial coefficients of 6D\ SCFTs. We have presented strong
evidence that for 6D\ SCFTs, there are actually large families of such
functions, and have delineated the precise boundaries of this monotonic
region of ``m-space."  This is in stark
contrast to the case of lower-dimensional (i.e. $D\leq4$) systems, where there
is a unique quantity, even for SCFTs. We have also shown that the quantity
$a_{6D}$ is indeed monotonic for all known flows for 6D\ SCFTs. In the
remainder of this section we discuss some avenues of future investigation.

In this note our primary emphasis has been on the linear combinations of
anomaly polynomial coefficients which are monotonic under all supersymmetric
flows. It would be quite illuminating to express these coefficients in terms
of the conformal anomaly coefficients of a 6D\ CFT, perhaps using an extension
of the results in reference \cite{Cordova:2015fha},
or perhaps using the methods of reference \cite{EHOLOG}.

Finally, in some sense, the classification of 6D\ SCFTs provides significantly
more data than just a few pieces of numerical data. It would be very
interesting to map out the full set of RG\ flows for 6D\ SCFTs.

\section*{Acknowledgements}

We thank D. S. Park for several discussions and collaboration at an early
stage in this work. We especially thank C. Vafa for helpful discussions and
correspondence. We also thank C. Cordova and T. Dumitrescu, C. P. Herzog, D. R. Morrison, B. Wecht and K.
Yonekura for several helpful discussions. The work of TR is supported by NSF
grant PHY-1067976 and by the NSF GRF under DGE-1144152.

%%%%%%%%%%%%%%%%%%%%%%%%%%%%%%%%%%%%%%%%%%%%%%%%%%%%%%%%%%%%%%%%%%%%%%%%%%%%%%%%%%%%%%%%%%%%%%%%%%%%%%%%%%%

\appendix

\newpage

\bibliographystyle{utphys}
\bibliography{sixDatheorem}

\end{document}